\documentclass[10pt]{article}
\usepackage{amssymb,amsmath,amsthm,epsfig}
\usepackage{color}
\theoremstyle{plain}

\newtheorem{theorem}{Theorem}[section]
\newtheorem{corollary}{Corollary}[section]
\newtheorem{lemma}{Lemma}[section]

\newtheorem{definition}{Definition}[section]

\numberwithin{equation}{section}
\usepackage{graphicx}
\usepackage{epstopdf}

\begin{document}
\title{A note on traveling wave solutions to the two component
Camassa-Holm equation}
\author{K.Mohajer\\
Department of Mathematics and Statistics\\
University of Saskatchewan\\
106 Wiggins Road\\
Saskatoon, SK, S7N 5E6
CANADA\\
mohajer@math.usask.ca}
\date{\today}
\maketitle

\begin{abstract}
In this paper we show that non-smooth functions which are
distributional traveling wave solutions to the two component
Camassa-Holm equation are distributional traveling wave solutions to
the Camassa-Holm equation provided that the set $u^{-1}(c)$, where
$c$ is the speed of the wave, is of measure zero. In particular
there are no new peakon or cuspon solutions beyond those already
satisfying the Camassa-Holm equation. However, the two component
Camassa-Holm equation has distinct from Camassa-Holm equation smooth
traveling wave solutions as well as new distributional solutions
when the measure of $u^{-1}(c)$ is not zero. We provide examples of
such solutions.
\end{abstract}

The Camassa-Holm equation \cite{camassa-holm}
\begin{equation}\label{CH}
u_{t}+\kappa u_{x}-u_{xxt}+3uu_{x}=2u_{x}u_{xx}+uu_{xxx},
\end{equation}
arises as a model for the unidirectional propagation of shallow
water waves over a flat bottom, $u(x,t)$ representing the water's
free surface, and $\kappa\in\mathbb{R}$ being a parameter related to
the critical shallow water speed. Camassa and Holm
\cite{camassa-holm} discovered that the equation has non-smooth
solitary waves that retain their individual characteristics through
the interaction and eventually emerge with their original shapes and
speeds. The traveling wave solutions of the Camassa-Holm equation
have been classified by J. Lenells \cite{lenells}. An alternative,
and useful for generalizations form of this equation is
\begin{equation}
m_{t}+um_{x}+2mu_{x}=0,
\end{equation}
where $m=u-u_{xx}+\frac{1}{2}\kappa$.\\ One such a generalization
has been introduced by M. Chen, S. Liu and Y. Zhang \cite{liu}:
\begin{equation}\label{CH2}
\begin{cases}
m_{t}+um_{x}+2mu_{x}-\rho\rho_{x}=0,\\
\rho_{t}+(\rho u)_{x}=0,
\end{cases}
\end{equation}
The traveling wave solutions are obtained by setting  $u=u(x-ct)$
and $\rho=\rho(x-ct)$. In this case, easy manipulations show that
\eqref{CH2} can be written as follows
\begin{equation}\label{travCH2}
\begin{cases}
-2c(u^{\prime}-u^{\prime\prime\prime})+2\kappa
u^{\prime}+3(u^2)^{\prime}+((u^{\prime})^2)^{\prime}-(u^2)^{\prime\prime\prime}=(\rho^2)^{\prime},\\
-c\rho^{\prime}+(\rho u)^{\prime}=0.
\end{cases}
\end{equation}
These equations are valid in the sense of distributions, if $u\in
H^{1}_{loc}(\mathbb{R})$ and $\rho\in L^{2}_{loc}(\mathbb{R})$.
Indeed, for a given function $\rho$, if
$(\rho^2)^{\prime}\in\mathcal{D}^{\prime}(\mathbb{R})$, then
$\rho\in L^{2}_{loc}(\mathbb{R})$. \\Since every distribution has a
primitive which is a distribution (see \cite{gelfand}), we can
integrate and then rewrite
\begin{equation}\label{CH2sol}
\begin{cases}
(v^2)^{\prime\prime}=(v^{\prime})^{2}+p(v)-\rho^2,\\
\rho v=B_1.
\end{cases}
\end{equation}
where $v=u-c$ and $p(v)=3v^2+(2\kappa+4c)v+K$ for some constants $K$
and $B_1$.
\begin{definition}\label{dist-trav-def}
A pair of functions $(u,\rho)$ where $u\in H^{1}_{loc}(\mathbb{R})$
and $\rho\in L^{2}_{loc}(\mathbb{R})$, is called a traveling wave
solution for \eqref{CH2} if $u$ and $\rho$ satisfy \eqref{CH2sol} in
the sense of distributions.
\end{definition}
The following Lemma is due to J. Lenells \cite{lenells}.
\begin{lemma}\label{lenells}
Let $p(v)$ be a polynomial with real coefficient. Assume that $v\in
H^{1}_{loc}(\mathbb{R})$ satisfies
\begin{equation}
(v^2)^{\prime\prime}=(v^{\prime})^2+p(v)\ \ \ \text{in\ \
$\mathcal{D}^{\prime}(\mathbb{R})$}.
\end{equation}
Then
\begin{equation}
v^k\in C^j(\mathbb{R})\ \ \ \text{for\ \ $k\geq 2^j$}.
\end{equation}
\end{lemma}
In our case, we have the following generalization:
\begin{lemma}\label{vkCj}
Let $p(v)$ be a polynomial with real coefficients. Assume that $v\in
H^{1}_{loc}(\mathbb{R})$ and $\rho\in L^{2}_{loc}(\mathbb{R})$
satisfy the following system in
$\mathcal{D}^{\prime}\bigl(\mathbb{R}\bigr)$:
\begin{equation}\label{disE}
\begin{cases}
(v^2)^{\prime\prime}=(v^{\prime})^2+p(v)-\rho^2,\\
\rho v=B_1.
\end{cases}
\end{equation}
Then
\begin{equation}
v^k\in C^j\bigl(\mathbb{R}\bigr)\ \ \ \text{for\ \ $k\geq 2^j$\ \
and\ \ $j\geq 0$}.
\end{equation}
\end{lemma}
\begin{proof}
Since $v\in H^{1}_{loc}(\mathbb{R})$ and $\rho\in
L^{2}_{loc}(\mathbb{R})$, \eqref{disE} implies that
$(v^2)^{\prime\prime}\in L^{1}_{loc}(\mathbb{R})$. Therefore,
$(v^2)^{\prime}$ is absolutely continuous and $v^2\in
C^{1}(\mathbb{R})$. Also, since $v\in H^{1}_{loc}(\mathbb{R})$, then
$v$ is absolutely continuous and we can claim
\[
(v^k)^{\prime}=\frac{k}{2}\bigl(v^{k-2}(v^2)^{\prime}\bigr)\ \ \
\text{for\ \ $k\geq 3$}.
\]
To see why the claim is true, we first note that in fact, it is
obviously true if $k$ is an even number. Also, note that since the
first derivative of an absolutely continuous function exists almost
everywhere, in taking the first derivative of the product of two
absolutely continuous functions we can use the Leibniz Rule almost
everywhere. Now, if $k$ is an odd number, let's say $k=2n+1$, then
we can write
\begin{equation*}
\begin{split}
(v^k)^{\prime}&=(v^{2n}v)^{\prime}=v(v^{2n})^{\prime}+v^{\prime}v^{2n}\\
&=v(nv^{2(n-1)})(v^2)^{\prime}+\frac{1}{2}(v^2)^{\prime}v^{2n-1}\\
&=\frac{k}{2}v^{k-2}(v^2)^{\prime}.
\end{split}
\end{equation*}
Thus, we have
\begin{equation*}
\begin{split}
(v^k)^{\prime\prime}&=\frac{k}{2}\bigl(v^{k-2}(v^2)^{\prime}\bigr)^{\prime}\\
&=\frac{k}{2}\bigl((v^{k-2})^{\prime}(v^2)^{\prime}+v^{k-2}(v^2)^{\prime\prime}\bigr)\\
&=k(k-2)v^{k-2}(v^{\prime})^2+\frac{k}{2}v^{k-2}(v^2)^{\prime\prime}\
\ \ \text{for\ \ $k\geq 3$}.
\end{split}
\end{equation*}
Substituting from \eqref{disE} we have
\begin{equation}\label{vkdoubleprime}
\begin{split}
(v^k)^{\prime\prime}&=k(k-2)v^{k-2}(v^{\prime})^2+\frac{k}{2}v^{k-2}\bigl((v^{\prime})^2+p(v)-\rho^2\bigr)\\
&=k(k-\frac{3}{2})v^{k-2}(v^{\prime})^2+\frac{k}{2}v^{k-2}p(v)-\frac{k}{2}B_1v^{k-3}\rho.
\end{split}
\end{equation}
For $k\geq 3$ the right hand side of the above equation belongs to
$L^{1}_{loc}(\mathbb{R})$. Therefore
\begin{equation}
v^k\in C^1\bigl(\mathbb{R}\bigr)\ \ \ \text{for\ \ $k\geq 2$}.
\end{equation}
Thus, the assertion holds for $j=1$. We proceed by induction on $j$.
Suppose
\[
v^k\in C^{j-1}\bigl(\mathbb{R}\bigr)\ \ \ \text{for\ \ $k\geq
2^{j-1}$\ \ and\ \ $j\geq 2$}.
\]
Then for $k\geq 2^j$ we have
\begin{equation}
\begin{split}
v^{k-2}(v^{\prime})^2&=\frac{1}{2^{j-1}}(2^{j-1}v^{2^{j-1}-1}v^{\prime})\frac{1}{k-2^{j-1}}((k-2^{j-1})v^{k-2^{j-1}-1}v^{\prime})\\
&=\frac{1}{2^{j-1}(k-2^{j-1})}(v^{2^{j-1}})^{\prime}(v^{k-2^{j-1}})^{\prime}\in
C^{j-2}(\mathbb{R}).
\end{split}
\end{equation}
Also, we have $v^{k-2}p(v)\in C^{j-1}(\mathbb{R})$ and
$v^{k-3}\rho=B_1v^{k-4}\in C^{j-2}(\mathbb{R})$. Therefore the right
hand side of equation \eqref{vkdoubleprime} belongs to
$C^{j-2}(\mathbb{R})$. Hence,
\begin{equation*}
v^k\in C^j\bigl(\mathbb{R}\bigr)\ \ \ \text{for\ \ $k\geq 2^j$}.
\end{equation*}
\end{proof}
\textbf{Remark.} Lemma \eqref{vkCj} implies that $v^{\prime}$ is
possibly discontinuous only at points where $v=0$. In fact, a much
stronger result is true:
\begin{corollary}\label{vlocCinfinity}
If $v\in H^{1}_{loc}(\mathbb{R})$ and $\rho\in
L^{2}_{loc}(\mathbb{R})$ satisfy \eqref{disE} in
$\mathcal{D}^{\prime}\bigl(\mathbb{R}\bigr)$, then
\[
v\in C^{\infty}\bigl(\mathbb{R}\setminus v^{-1}(0)\bigr)
\]
and
\[
\rho\in C^{\infty}\bigl(\mathbb{R}\setminus v^{-1}(0)\bigr).
\]
\end{corollary}
\begin{proof}
Suppose $k\geq 2$. Then $v^k\in C^1(\mathbb{R})$. Therefore
\[
kv^{k-1}v^{\prime}=(v^k)^{\prime}\in C(\mathbb{R}).
\]
This implies that $v^{\prime}\in C\bigl(\mathbb{R}\setminus
v^{-1}(0)\bigr)$. Thus, $v\in C^{1}\bigl(\mathbb{R}\setminus
v^{-1}(0)\bigr)$.\\ Now, assume that $v\in
C^{j}\bigl(\mathbb{R}\setminus v^{-1}(0)\bigr)$ for $j\geq 1$. For
$k\geq 2^{j+1}$, we have $v^k\in C^{j+1}(\mathbb{R})$. Therefore
\[
kv^{k-1}v^{\prime}=(v^k)^{\prime}\in C^{j}(\mathbb{R}).
\]
This shows that $v^{\prime}\in C^{j}\bigl(\mathbb{R}\setminus
v^{-1}(0)\bigr)$. Hence, $v\in C^{j+1}\bigl(\mathbb{R}\setminus
v^{-1}(0)\bigr)$. Thus, $u$ is in the desired space. Now the
statement for $\rho$ follows from the second equation of
\eqref{CH2sol}.
\end{proof}
\textbf{Remark.} Since $v=u-c$, Corollary \eqref{vlocCinfinity}
shows that $u\in C^{\infty}\bigl(\mathbb{R}\setminus
u^{-1}(c)\bigr)$.\\

Since $\mathbb{R}\setminus u^{-1}(c)$ is an open set, we have
\[
\mathbb{R}\setminus u^{-1}(c)=\bigcup_{i=1}^{\infty}(a_i,b_i).
\]
So, u is smooth in every interval $(a_i,b_i)$ where the following
Lemma holds (below $(a_i,b_i)=(a,b)$):
\begin{lemma}
Let $(u,\rho)$ be a traveling wave solution to \eqref{CH2}. Suppose
$u$ is smooth in the interval $(a,b)$. Then in the interval $(a,b)$,
$u$ satisfies the following equation:
\begin{equation}\label{polyCH2}
(u-c)^{2}u^{\prime2}=P(u),
\end{equation}
where
\begin{equation}\label{upoly}
P(u)=(u^2+\kappa u+A)(u-c)^2+C(u-c)+B,
\end{equation}
and $A$, $B$ and $C$ are some constants.
\end{lemma}
\begin{proof}
Since both $u$ and $\rho$ are smooth in $(a,b)$ we use standard
calculus rules. By the first equation of \eqref{CH2sol}, we have
\begin{equation*}
2(v^{\prime})^2+2vv^{\prime\prime}=(v^{\prime})^2+p(v)-\rho^2.
\end{equation*}
Therefore,
\begin{equation*}
(v^{\prime})^2+2vv^{\prime\prime}=p(v)-\rho^2.
\end{equation*}
Multiplying by $v^{\prime}$ we have
\[
(v^{\prime})^3+v\bigl((v^{\prime})^2\bigr)^{\prime}=v^{\prime}p(v)-v^{\prime}\rho^2.
\]
Thus,
\[
\bigl(v(v^{\prime})^2\bigr)^{\prime}=v^{\prime}p(v)-v^{\prime}\rho^2.
\]
Hence,
\[
\bigl(v(v^{\prime})^2\bigr)^{\prime}=(3v^2+(2\kappa+4c)v+K)v^{\prime}-\frac{Bv^{\prime}}{v^2},
\]
where $B=B_{1}^2$.\\
Integration yields
\begin{equation*}
v(v^{\prime})^2=v^3+(\kappa+2c)v^2+Kv+\frac{B}{v}+C.
\end{equation*}
Now, multiplying this equation by $v$ we get
\begin{equation*}
v^2(v^{\prime})^2=\bigl(v^2+(\kappa+2c)v+K\bigr)v^2+Cv+B.
\end{equation*}
Substituting $v=u-c$ and simplifying, we have
\[
(u-c)^2(u^{\prime})^2=(u^2+\kappa u+A)(u-c)^2+C(u-c)+B,
\]
for some constant A.
\end{proof}
\begin{theorem}
Suppose $(u,\rho)$ is a non-smooth traveling wave solution to
\eqref{CH2}. If $u^{-1}(c)$ is a set of measure zero, then $u$ is a
solution to the Camassa-Holm equation.
\end{theorem}
\begin{proof}
Suppose $\xi\in\mathbb{R}\setminus u^{-1}(c)$. Since, $u^{-1}(c)\neq
\varnothing$, there exists an $\eta\in u^{-1}(c)$ such that either
$\xi>\eta$ or $\xi<\eta$. Without loss of generality, assume that
$\xi<\eta$. Let $\eta_0=\inf\lbrace\eta\in
u^{-1}(c):\eta>\xi\rbrace$. Since $u^{-1}(c)$ is a closed set,
$\eta_0\in u^{-1}(c)$. So, $(\xi,\eta_0)\subseteq
\mathbb{R}\setminus u^{-1}(c)$. Thus, we have proved that there
exists an $\eta\in u^{-1}(c)$ such that either
$(\xi,\eta)\subseteq\mathbb{R}\setminus u^{-1}(c)$ or
$(\eta,\xi)\subseteq\mathbb{R}\setminus u^{-1}(c)$. Now, consider
the equation \eqref{polyCH2} and set $F(u)=\frac{P(u)}{(u-c)^2}$. We
claim that $B$ in \eqref{polyCH2} equals $0$. Suppose $B\neq 0$.
Since $B=B_1^2$, we have $B > 0$. Then \eqref{upoly} implies that
\[
\frac{1}{\sqrt{F(u)}}=\frac{1}{\sqrt{B}}\lvert u-c \rvert +
\mathcal{O}\bigl((u-c)^2\bigr)\ \ \ \ \ u\rightarrow c.
\]
On the other hand, we have
\[
\frac{d\xi}{du}=\pm\frac{1}{\sqrt{F(u)}}.
\]
Since $u\in C(\mathbb{R})$, for $\xi$ close enough to $\eta$,
integration yields
\begin{equation}
\lvert \xi-\eta
\rvert=\frac{1}{2\sqrt{B}}(u-c)^2+\mathcal{O}\bigl((u-c)^3\bigr)\ \
\ \ \ u\rightarrow c.
\end{equation}

Therefore,
\[
\lvert \xi-\eta
\rvert=\frac{1}{2\sqrt{B}}(u-c)^2\bigl(1+\mathcal{O}\bigl(u-c)\bigr)\
\ \ \ \ u\rightarrow c.
\]
So,
\[
\lvert \xi-\eta
\rvert^{\frac{1}{2}}=\frac{1}{\sqrt{2\sqrt{B}}}\lvert u-c \rvert
\sqrt{\bigl(1+\mathcal{O}\bigl(u-c)\bigr)}\ \ \ \ \ u\rightarrow c.
\]
Thus,
\[
\lvert \xi-\eta
\rvert^{\frac{1}{2}}=\frac{1}{\sqrt{2\sqrt{B}}}\lvert u-c \rvert
\bigl(1+\mathcal{O}\bigl(u-c)\bigr)\ \ \ \ \ u\rightarrow c.
\]
Hence,
\[
\lvert \xi-\eta
\rvert^{\frac{1}{2}}=\frac{1}{\sqrt{2\sqrt{B}}}\lvert u-c \rvert
+\mathcal{O}\bigl((u-c)^2\bigr)\ \ \ \ \ u\rightarrow c.
\]
This implies that
\[
(u-c)=\mathcal{O}\bigl((\xi-\eta)^{\frac{1}{2}}\bigr)\ \ \ \ \
\xi\rightarrow \eta.
\]
Therefore,
\[
(u-c)^2=\mathcal{O}(\xi-\eta)\ \ \ \ \ \xi\rightarrow \eta.
\]
Thus, we have
\begin{equation}
\lvert u-c \rvert=\sqrt{2\sqrt{B}}\lvert \xi-\eta
\rvert^{\frac{1}{2}}+\mathcal{O}(\xi-\eta)\ \ \ \ \ \xi\rightarrow
\eta.
\end{equation}

Hence,
\begin{equation*}
\begin{split}
\lvert \xi-\eta \rvert^{-\frac{1}{2}}-\sqrt{2\sqrt{B}}\lvert u-c
\rvert^{-1} &= \mathcal{O}\bigl(\frac{\lvert \xi-\eta \rvert^{\frac{1}{2}}}{u-c}\bigr)\\
&=\mathcal{O}(1)\ \ \ \ \ \xi\rightarrow\eta.
\end{split}
\end{equation*}
So,
\begin{equation}\label{reciprocal_estimate}
\lvert u-c \rvert^{-1}=\frac{1}{\sqrt{2\sqrt{B}}}\lvert \xi-\eta
\rvert^{-\frac{1}{2}}+\mathcal{O}(1)\ \ \ \ \ \xi\rightarrow \eta.
\end{equation}
On the other hand, from \eqref{upoly} we have
\begin{equation}\label{u_prime_estimate1}
\lvert u^{\prime} \rvert=\sqrt{B}(u-c)^{-1}+\mathcal{O}(1)\ \ \ \ \
\xi\rightarrow\eta.
\end{equation}
Now combining \eqref{u_prime_estimate1} and
\eqref{reciprocal_estimate}, we have
\begin{equation}
\lvert u^{\prime}\rvert=\frac{\sqrt[4]{B}}{\sqrt{2}}\lvert \xi-\eta
\rvert ^{-\frac{1}{2}}+\mathcal{O}(1)\ \ \ \ \ \xi\rightarrow \eta.
\end{equation}
Hence, $u^{\prime}\notin L_{loc}^2(\mathbb{R})$. This contradiction
shows that $B=0$. Therefore, the second equation of \eqref{CH2sol}
implies that $\rho=0$ almost everywhere.
\end{proof}

Now, we provide an example of a smooth solution of \eqref{CH2} that
is not a solution of Camassa-Holm equation.

\bigskip
\textbf{Example.} Let $P(u)$ be as in the previous Theorem. Observe
that $P(u)=(u-G)^2(u-L)^2$ if and only if
\begin{equation}
\begin{cases}
\kappa=2\bigl(c-(L+G)\bigr),\\
A=2c\kappa-c^2+(L+G)^2+2LG,\\
C=2cA-\kappa c^2-2LG(L+G),\\
B=Cc-Ac^2+L^2G^2.
\end{cases}
\end{equation}
Suppose $\lvert u \rvert<1$ and $c>1$. Therefore, if $G=-1$ and
$L=1$, integration yields
\begin{equation}
(1-u)^{1-c}(1+u)^{1+c}=e^{2(\xi-\xi_0)}.
\end{equation}

\begin{figure}[!h]
\centering
\includegraphics[scale=.2]{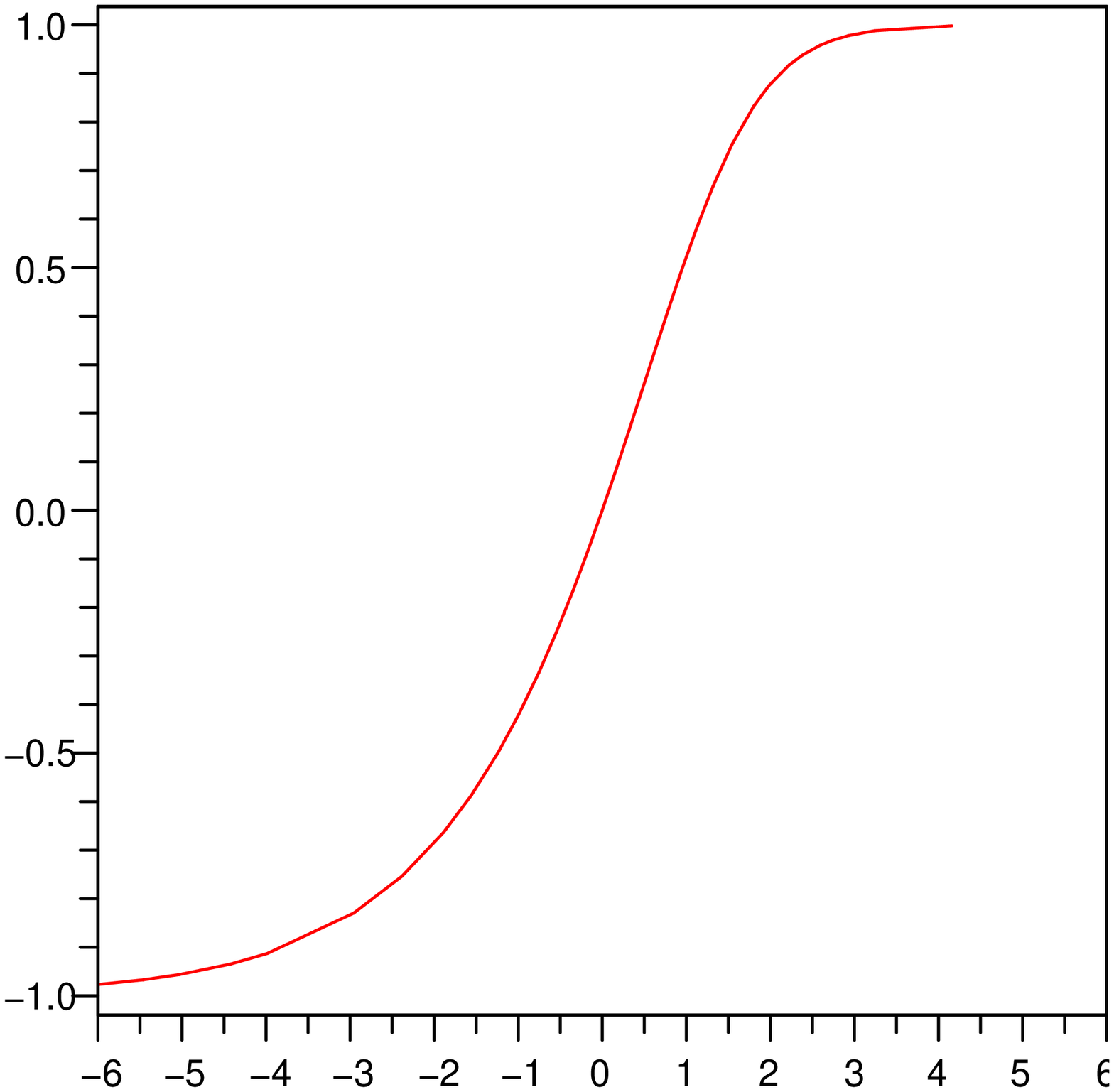}\ \ \ \ \
\includegraphics[scale=.2]{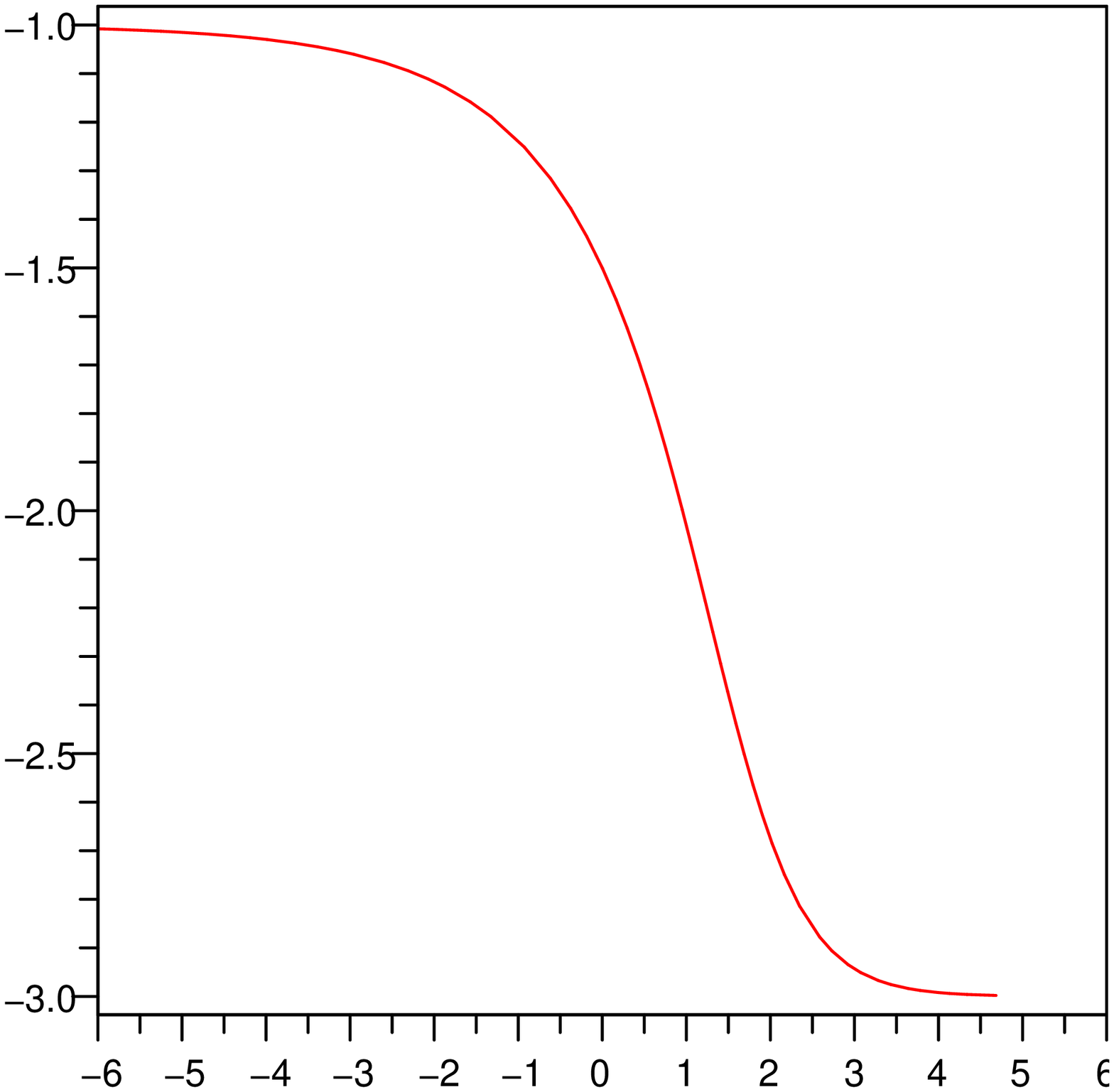}
\caption{($u$ on the left and $\rho$ on the right) A smooth solution
of \eqref{CH2} which is not a solution of Camassa-Holm equation.}
\label{smoothCH2}
\end{figure}

Let's say $c=2$ and $\xi_0=0$. We observe that the equation
\[
\frac{(1+u)^3}{1-u}=e^{2\xi},
\]
provides a smooth solution of \eqref{CH2} which is not a solution of
Camassa-Holm equation. See figure \ref{smoothCH2}.

\bigskip
The following Lemma provides necessary and sufficient conditions for
a piecewise smooth function to be a distributional solution to
\eqref{CH2}.

\bigskip
\begin{lemma}
Suppose $u$ is a piecewise smooth function. The pair $(u,\rho)$ is a
distributional solution to \eqref{CH2} in the sense of definition
\ref{dist-trav-def} if and only if all of the
following conditions hold:\\
1. $u\in H^1_{loc}(\mathbb{R})$ and $\rho\in
L^2_{loc}(\mathbb{R})$.\\
2. $(u-c)^2 \in W^{2,1}_{loc}(\mathbb{R})$.\\
3. $u$ and $\rho$ satisfy the equation \eqref{CH2sol} locally with
the same constant $K$ on every interval where $u$ is smooth.
\end{lemma}
\begin{proof}
The part $(\Rightarrow)$ is easy. For the converse $(\Leftarrow)$,
we note that since $(u-c)^2 \in W^{2,1}_{loc}(\mathbb{R})$, then
$((u-c)^2)^{\prime}$ is absolutely continuous and has no jumps.
Therefore, $((u-c)^2)^{\prime\prime}$ defines a regular distribution
\cite{gelfand} .Thus, every term in the equation \eqref{CH2sol} can
be represented by an integral that defines a distribution on the
space of test functions and we are allowed to write each integral as
a finite sum of integrals over local intervals and use condition 3
to prove that $u$ and $\rho$ satisfy \eqref{CH2sol} in the sense of
distributions.
\end{proof}

\bigskip

\textbf{Remark.} We note that if the measure of $u^{-1}(c)$ is not
zero, then the equation \eqref{CH2sol} implies that $\rho^2=K$ on
$u^{-1}(c)$. However in the Camassa-Holm equation if the measure of
$u^{-1}(c)$ is not zero, then $K=0$ because $\rho=0$.
This implies that solutions of the form given in the following example cannot arise from the Camassa-Holm equation.\\

\bigskip

\begin{figure}[!h]
\centering
\includegraphics[scale=.25]{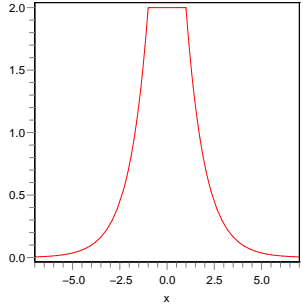}
\caption{$u(x)$ is a solution of \eqref{CH2} but it is not a
solution of Camassa-Holm equation.} \label{stumpeakon}
\end{figure}

\textbf{Example.} Set $\kappa=0$. The pair of functions $(u,\rho)$
given by
\begin{equation*}
u(x)=
\begin{cases}
ce^{1-\lvert x \rvert}\ \ \ \text{if}\ \ \lvert x \rvert>1,\\
c\ \ \ \ \ \ \ \ \ \ \,\text{if}\ \ \lvert x \rvert<1,
\end{cases}
\end{equation*}
\begin{equation*}
\rho(x)=
\begin{cases}
c\ \ \ \text{if}\ \ \lvert x \rvert<1,\\
0\ \ \ \text{if}\ \ \lvert x \rvert>1,
\end{cases}
\end{equation*}
is a solution to \eqref{CH2} but $u$ is not a solution of
Camassa-Holm equation. To see this, observe that the left hand side
derivative of $u$ at $-1$ and the right hand side derivative of $u$
at $1$ are non-zero and finite in contrast with the Camassa-Holm
equation for which Lenells \cite{lenells} showed that if the measure
of $u^{-1}(c)$ is not zero, then these limits cannot be finite. See
figure \ref{stumpeakon}.
\bigskip
\bigskip
\vspace{2cm}
\begin{definition}\label{defpeakon}
Suppose $f$ is a continuous function on $\mathbb{R}$.\\
1. We say $f$ has a peak at $x$ if $f$ is smooth locally on both
sides of $x$ and
\[
0\neq\lim_{y\downarrow x}f^{\prime}(y)=-\lim_{y\uparrow
x}f^{\prime}(y)\neq \pm\infty.
\]
Traveling wave solutions of \eqref{CH2} with peaks are called
peakons.\\
2. We say $f$ has a cusp at $x$ if $f$ is smooth locally on both
sides of $x$ and
\[
\lim_{y\downarrow x}f^{\prime}(y)=-\lim_{y\uparrow x}f^{\prime}(y)=
\pm\infty.
\]
Traveling wave solutions of \eqref{CH2} with cusps are called
cuspons.\\
3. We say that $f$ has a stump if there is an interval $[a,b]$ on
which $f$ is a constant and $f$ is smooth locally to the left of $a$
and to the right of $b$ and
\[
0\neq\lim_{x\uparrow a}f^{\prime}(x)=-\lim_{x\downarrow
b}f^{\prime}(x).
\]
Traveling wave solutions of \eqref{CH2} with stumps are called
stumpons. Note that, in the definition of a stump the limits can be
either finite or infinite.
\end{definition}
\bigskip
Theorem 0.1 limits the existence of new distributional peakon or
cuspon solutions to the \eqref{CH2}.
\bigskip
\begin{corollary}
Every peakon or cuspon traveling wave solution to \eqref{CH2} is a
traveling wave solution to the Camassa-Holm equation.
\end{corollary}

\bigskip

\begin{figure}[!h]
\centering
\includegraphics[scale=.25]{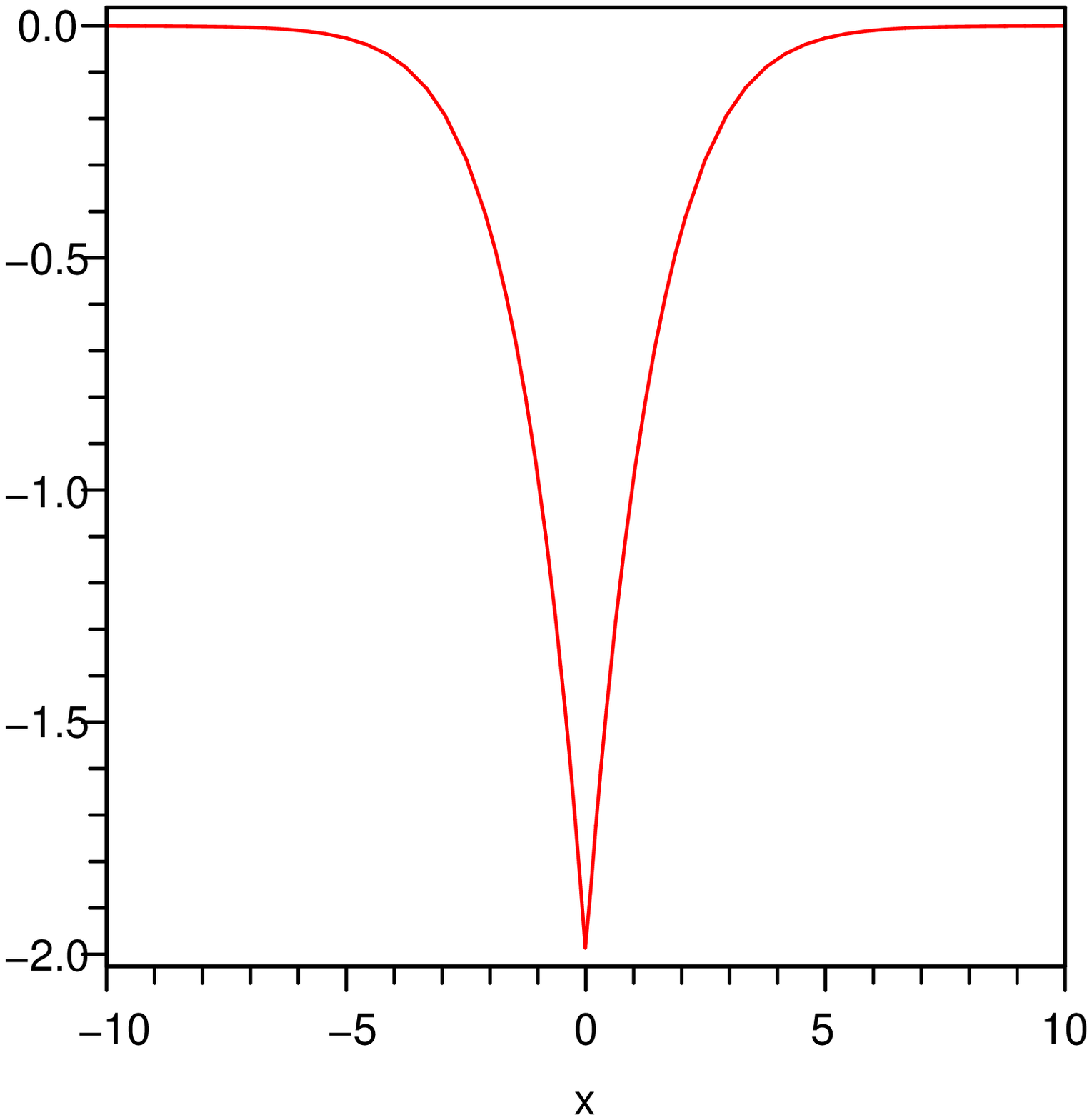}\ \ \
\includegraphics[scale=.25]{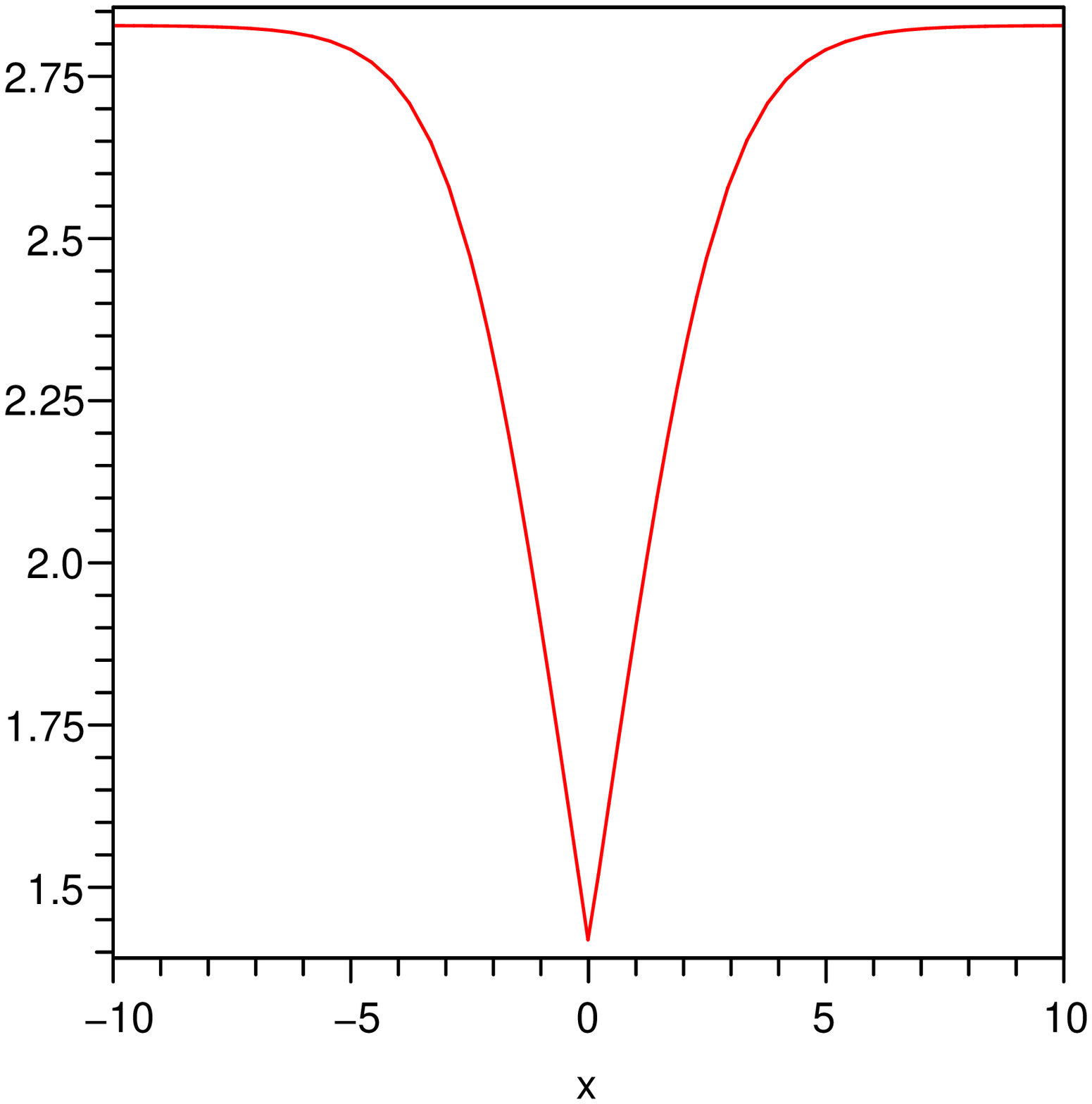}
\caption{($u$ on the left and $\rho$ on the right) This pair is not
a distributional traveling wave solution of \eqref{CH2}. $c=2$ and
$K_1=-1$.}\label{notsolution}
\end{figure}

Finally we would like to comment on the peaked solution reported in
\cite{liu}. For reasons explained below, that solution is not a
distributional solution. First, we note that by Corollary
\eqref{vlocCinfinity} the non-smooth points of a distributional
solution $u$ can only appear when $u=c$. Also, Lemma \eqref{vkCj}
shows that if $(u,\rho)$ is a traveling wave solution to
\eqref{CH2}, then $(u-c)^2\in C^1(\mathbb{R})$. Now, consider the
peaked function (see \cite{liu})
\[
u=\chi+\sqrt{\chi^2-c^2},\ \
\rho=\sqrt{-cK_1}\biggl(1+\sqrt{\frac{\chi+c}{\chi-c}}\biggr),\ \
\chi=-(c+K_1)\cosh(x-ct)+K_1,
\]

where $K_1=-\frac{1}{4}\kappa$, $K_1<0$ and $ c >\lvert K_1
\rvert>0$. Away from it's non-smooth point, $u$ is a solution to
\eqref{CH2}. However, It is clear that $u$ is not smooth at $\xi=0$
and $u(0)=-c$. Furthermore, $(u-c)^2\notin C^1(\mathbb{R})$ because
\[
\lim_{\xi\downarrow 0}
\bigl((u-c)^2\bigr)^{\prime}-\lim_{\xi\uparrow 0}
\bigl((u-c)^2\bigr)^{\prime}=-8c\sqrt{c(c+K_1)},
\]
where $\xi=x-ct$. Therefore, $u$ is not a distributional solution to
\eqref{CH2} even though it superficially looks like a peakon
solution (see figure \ref{notsolution}).

\textbf{Acknowledgment.} I would like to thank professor J.
Szmigielski for suggesting the problem and tremendously helpful
comments.

\end{document}